\documentclass[]{seismica}

\usepackage{booktabs}
\usepackage{multirow}
\usepackage{array}
\usepackage{longtable}
\usepackage{xr}
\usepackage{xltabular}
\usepackage{pdfpages}

\nolinenumbers

\makeatletter
\newcommand*{\addFileDependency}[1]{
\typeout{(#1)}
\@addtofilelist{#1}
\IfFileExists{#1}{}{\typeout{No file #1.}}
}\makeatother
\newcommand*{\myexternaldocument}[1]{%
\externaldocument{#1}%
\addFileDependency{#1.tex}%
\addFileDependency{#1.aux}%
}
\myexternaldocument{main_supplement}

\title{A global-scale database of seismic phases from cloud-based picking at petabyte scale}

\reporttype{Data Report}

\author[1]{Yiyu Ni
	\orcid{0000-0001-5181-9700}
	\thanks{Corresponding author: niyiyu@uw.edu}
}
\author[1]{Marine A. Denolle
	\orcid{0000-0002-1610-2250}
}
\author[2]{Amanda M. Thomas
	\orcid{0000-0001-6997-3140}
}
\author[3]{Alex Hamilton
	\orcid{0009-0000-2377-4111}
}
\author[4]{Jannes Münchmeyer
	\orcid{0000-0002-4006-9673}
}
\author[5]{Yinzhi Wang
	\orcid{0000-0001-8505-0223}
}
\author[6]{Loïc Bachelot
	\orcid{0000-0003-0711-8581}
}
\author[3]{Chad Trabant
	\orcid{0000-0002-6184-3568}
}
\author[3]{David Mencin
	\orcid{0000-0001-9984-6724}
}

\affil[1]{Department of Earth and Space Sciences, University of Washington, Seattle, WA, USA}
\affil[2]{Department of Earth and Planetary Sciences, University of California, Davis, CA, USA}
\affil[3]{EarthScope Consortium, Washington, DC, USA}
\affil[4]{Université Grenoble Alpes, Université Savoie Mont Blanc, CNRS, IRD, Université Gustave Eiffel, ISTerre, Grenoble, France}
\affil[5]{Texas Advanced Computing Center, University of Texas, Austin, TX, USA}
\affil[6]{Cascadia Region Earthquake Science Center, University of Oregon, Eugene, OR, USA}

\credit{Conceptualization}{MD, AMT, JM, YW, YN, LB}
\credit{Methodology}{MD, JM, YW, YN, AMT}
\credit{Software}{YN, JM, YW}
\credit{Validation}{YN, AMT}
\credit{Formal Analysis}{YN, JM, YW, AMT}
\credit{Investigation}{YN, JM, YW, MD, AMT}
\credit{Resources}{MD, AMT}
\credit{Writing - Original draft}{AMT, MD, YN}
\credit{Writing - Review \& Editing}{AMT, MD, YN, LB, JM, CT, DM}
\credit{Visualization}{MD, YN}
\credit{Supervision}{MD, AMT}
\credit{Project Administration}{MD, AMT}
\credit{Funding Acquisition}{MD, YW}

\begin{document}

\makeseistitle{
\begin{summary}{Abstract}
We present the first global-scale database of 4.3 billion P- and S-wave picks extracted from 1.3 PB continuous seismic data via a cloud-native workflow. Using cloud computing services on Amazon Web Services, we launched $\sim$145,000 containerized jobs on continuous records from 47,354 stations spanning 2002–2025, completing in under three days. Phase arrivals were identified with a deep learning model, PhaseNet, through an open-source Python ecosystem for deep learning, SeisBench. To visualize and gain a global understanding of these picks, we present preliminary results about pick time series revealing Omori-law aftershock decay, seasonal variations linked to noise levels, and dense regional coverage that will enhance earthquake catalogs and machine-learning datasets. We provide all picks in a publicly queryable database, providing a powerful resource for researchers studying seismicity around the world. This report provides insights into the database and the underlying workflow, demonstrating the feasibility of petabyte-scale seismic data mining on the cloud and of providing intelligent data products to the community in an automated manner.
\end{summary}
}

\section{Introduction}
Detecting earthquakes by picking P- and S-wave arrival times is fundamental to seismology. It enables rapid estimation of earthquake source properties, provides early warning for potential ground shaking, and supports seismic hazard assessment. Picking the arrival time is also the first step in building earthquake catalogs. Traditional earthquake detection methods are typically unsupervised and rely on signal characteristics such as impulsivity, using techniques like the short-term average/long-term average (STA/LTA) filter \citep[e.g.,][]{allen1982automatic} or kurtosis-based approaches \citep{hibert2014automated}. However, these methods are highly sensitive to background seismic noise, limiting their effectiveness to small-magnitude events or recordings from particularly quiet stations.

Recent advances in seismic data processing techniques demonstrate that artificial intelligence (AI) and machine learning (ML) overcome these limitations and have shown high performance in earthquake detection \citep{perol2018convolutional, ross2018generalized, mousavi2019cred}, phase picking \citep{zhu2019phasenet, zhu2022earthquake, mousavi2020earthquake, ross2020p, michelini2021instance}, and phase association \citep{ross2019phaselink, mousavi2020earthquake, mcbrearty2023earthquake}. Supervised deep learning approaches to earthquake detection and phase identification require and have benefited from large, labeled datasets (e.g., several hundred thousand examples of P-waves, S-waves, and noise, along with the timing of body wave arrivals) for model training. These datasets are often compiled from analyst-reviewed phase picks cataloged by regional seismic networks and compiled by researchers for AI-readiness \citep{zhu2019phasenet, mousavi2019stanford, yeck2021leveraging, ni2023curated, zhu2025california}. Once trained, deep-learning-based detection and phase picking frameworks have significantly outperformed more conventional approaches in sensitivity and timing accuracy. They can detect arrivals in data with low signal-to-noise ratios and reliably pick arrivals to within less than 0.1~s \citep{zhu2019phasenet, mousavi2020earthquake}, without requiring manual intervention for hyperparameters. Additionally, once trained, these models are inference-only, making them extremely fast and scalable. These successes, combined with the abundance of readily available continuous seismic data, facilitate large-scale regional and global data mining efforts. 

Data volumes in seismology are expanding rapidly, with researchers now typically tackling datasets of TBs in size that require advanced computing strategies for analysis. Cloud computing is a promising infrastructure that can support this by enhancing the reliability, scalability, and accessibility of data \citep{gentemann2021science}, promoting reproducible and open science. Most importantly for users, cloud computing provides large-scale computing power close to the data archives, eliminating the need to download massive datasets and enabling users to run analyses in-place on the cloud using cloud-native tools (e.g., elastic computing, batch computing, scalable storage, and massive databases). The Southern California Seismic Network (SCEDC) has been hosting a copy of its entire data archive on the Amazon Web Services (AWS) Simple Storage Service (S3) since 2019 \citep{yu2021scedc}. The Northern California Earthquake Data Center (NCEDC) followed a similar model and architecture in 2024. The EarthScope Consortium is a non-profit organization that supports Earth science research by collecting, managing, and providing access to seismic and geodetic data collected worldwide from all United States (US) NSF-supported seismic experiments, a subset of the US regional seismic networks, and a selection of stations from global seismic networks \citep{zawacki2023advancing}. It serves the scientific community by operating and maintaining networks of instruments, curating data and metadata, and delivering these to end-users. In recent years, the EarthScope Consortium decided to migrate data collection, archiving, and delivery services to the cloud. 

Here we present results from one of the first large-scale, cloud-based seismic data-mining efforts. We processed approximately 1.3 PB of continuous seismic data from stations worldwide (see Figure~\ref{fig:1}). Leveraging modern cloud infrastructure, we developed a scalable cloud-native workflow to efficiently manage and analyze this extensive dataset. The sections below describe the waveform formats, cloud architecture, and compute resource utilization. We also describe the structure and contents of the resulting phase-pick database and provide public access through a web service. This work demonstrates the feasibility and advantages of using cloud computing for robust, high-throughput seismic waveform analysis. At the same time, the resulting database can serve as a starting point for the community to study seismicity globally. As in a typical catalog workflow, phase picking is the most runtime-intensive step. We believe that our database providing global-scale, high-quality phase picks has the potential to substantially accelerate seismicity studies worldwide.

\section{Methods}

\subsection{Data}
\begin{figure}[ht!]
    \includegraphics[width=\textwidth]{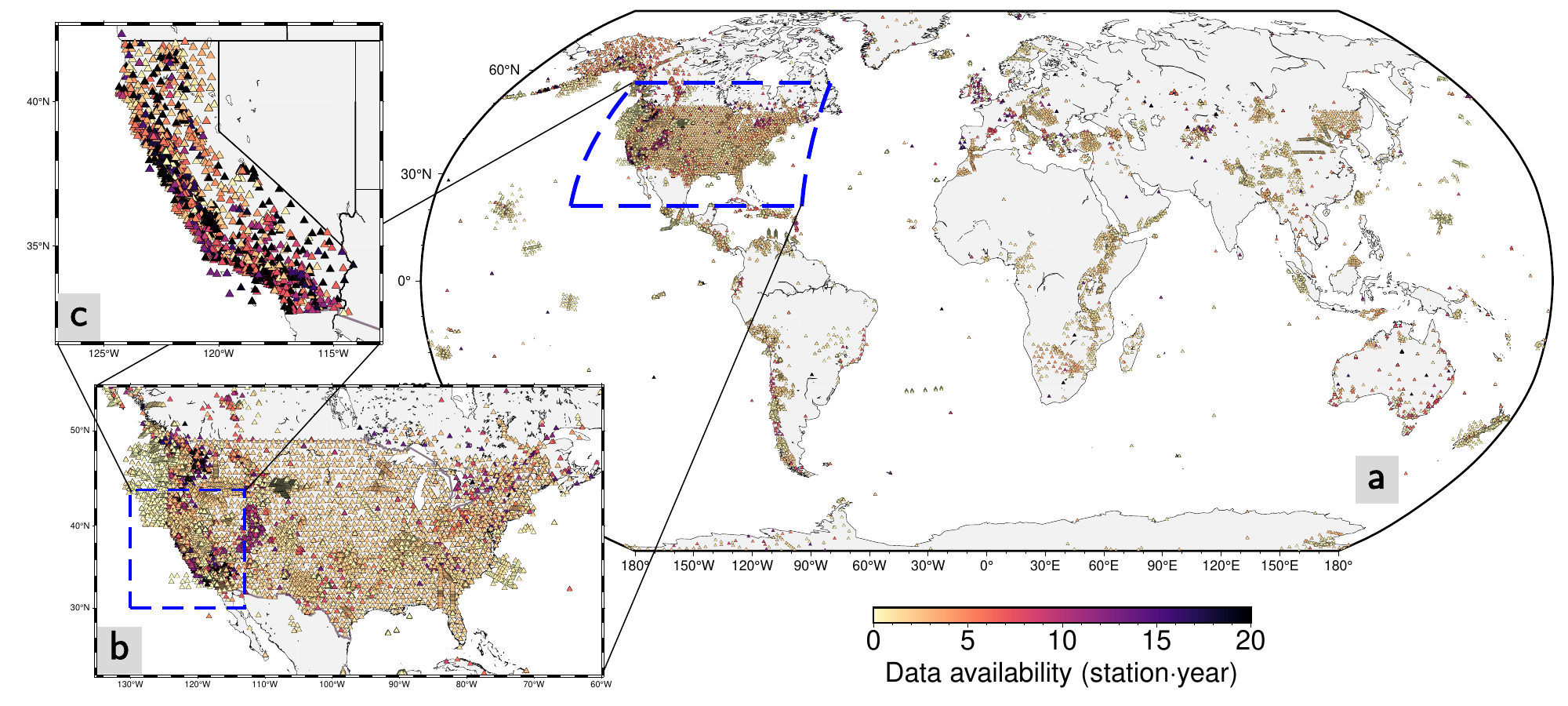} 
    \caption{Map of stations, displayed as triangles and color coded according to the Data availability in station-years included in EarthScope, NCEDC, and SCEDC archives. Panel (a) shows the distribution of global station data availability. Panel (b) shows a detailed view of stations in the United States, while Panel (c) shows data availability in California.}
    \label{fig:1}
\end{figure}

The NCEDC provides public access to continuous seismic waveform data through AWS, utilizing the AWS Open Data Sponsorship program. NCEDC continuous waveform data are hosted in the \texttt{ncedc-pds} S3 bucket in the us-east-2 region (Ohio, United States). Files are organized by network, year, and day of year, and are stored in miniSEED format, with each file representing one day of data from a single channel. Similarly, the SCEDC offers public access through the \texttt{scedc-pds} S3 bucket in the us-west-2 region (Oregon, United States). The EarthScope Consortium provides data access to credentialed users through an S3 Access Point, which gives fine-grained control over permissions and network access. While users interact with the Access Point much like a standard S3 bucket, internally, requests are routed through a Lambda function that handles S3 requests, enabling dynamic access management and custom logic. The combined archives represent 47,354 stations recording data between January 1, 2002, and March 31, 2025, for a total of more than 1.3 PB of continuous seismic data (Figure~\ref{fig:1}).
	
\subsection{Workflow}\label{sec:workflow}
\begin{figure}[ht!]
    \includegraphics[width=\textwidth]{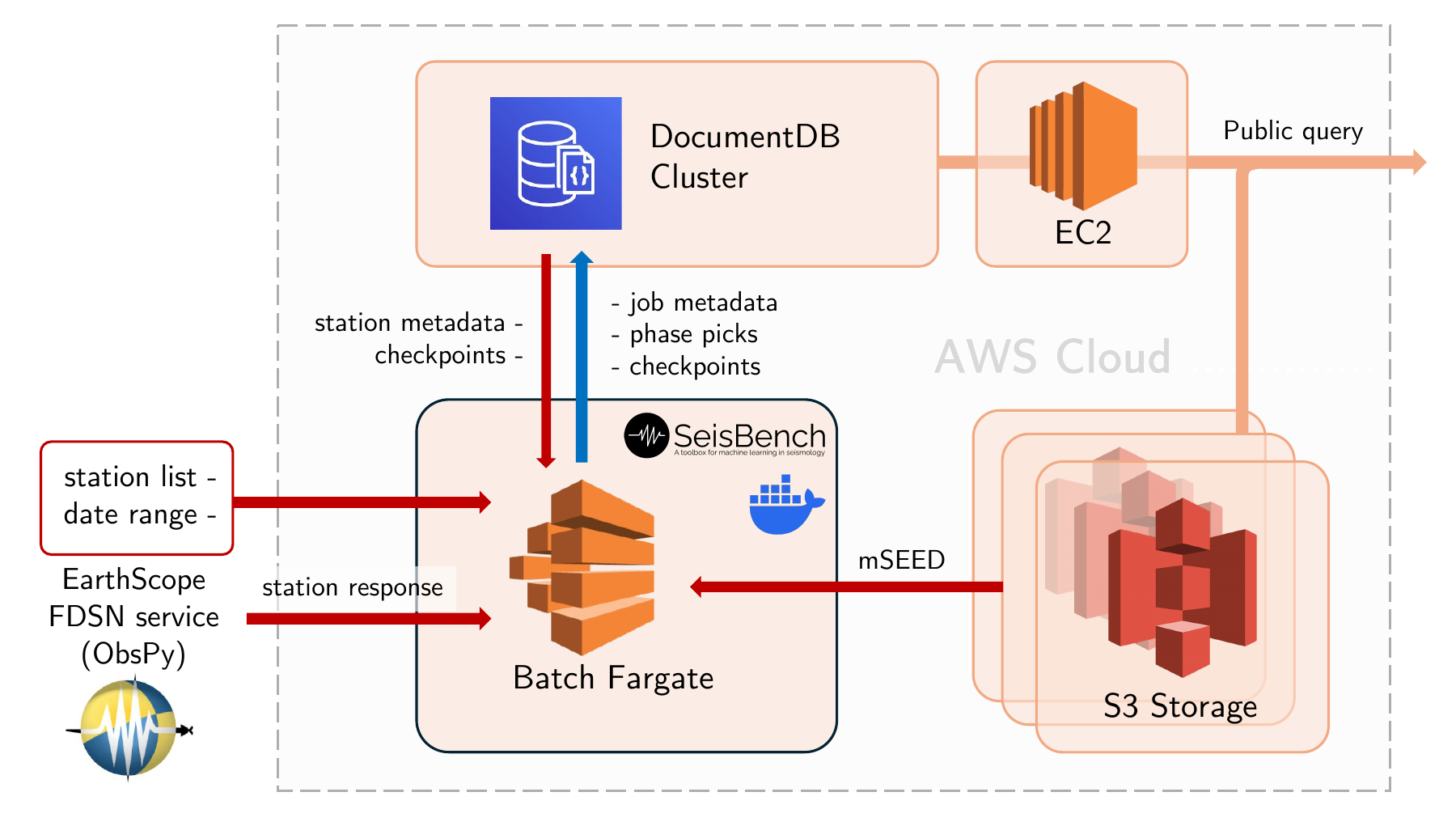} 
    \caption{The scalable cloud-native workflow for seismic phase picking. Containerized jobs are submitted to AWS Batch, which loads miniSEED seismic waveforms directly from AWS S3 buckets. Phase arrivals were identified with PhaseNet through the SeisBench implementation. A DocumentDB cluster is employed to store job metadata, picks, and checkpoints. Finally, an EC2 instance is used to provide a public database query service.}
    \label{fig:2}
\end{figure}

Several earthquake catalog building workflows exist \citep{walter2021easyquake, zhang2022loc, retailleau2022wrapper, sun2024deep}, some of which have focused on cloud deployments \citep{zhu2023quakeflow, krauss2023cloud}. Here, we develop a scalable cloud-native workflow for seismic data processing designed for large-scale data mining using AWS services, which we illustrate in Figure~\ref{fig:2}.

The workflow begins with a user-specified list of station codes and a defined time range. A unique combination of 40 stations and a 20-day time window is paired as one job, which is submitted to an AWS Batch computing queue. We request 8 vCPU and 16 GB RAM for each job, with interruptible Spot instances enabled. In contrast to the on-demand counterpart, Spot instances utilize unused AWS capacity with a discount (up to 90\%) but can be arbitrarily recalled. With appropriate retrial and checkpoint mechanisms implemented, Spot instances effectively optimize the cost efficiency of our workflow. The submitted jobs stay pending until requested resources are supplied, automatically elevating queued jobs into the running state until the account quota is reached. With a 12,000 vCPU account quota, the computing queue allows 1,500 jobs to run in parallel. A running job first retrieves instrument response information from the EarthScope International Federation of Digital Seismograph Stations (FDSN) \texttt{fdsnws-station} service via the ObsPy library \citep{beyreuther2010obspy}. A temporary user credential is requested from EarthScope to specifically enable EarthScope S3 access, while no credential is required for SCEDC and NCEDC. Waveforms are loaded directly from the S3 buckets by mapping the network code to the appropriate data center's bucket, bypassing the need for middleware returning waveforms such as the FDSN \texttt{fdsnws-dataselect} service. For this project, we processed all available seismic data from January 1, 2002, to March 31, 2025, across channels with the following codes: EH?, HH?, BH?, HN?, EP?, DP?, EL?, SL?, SH?, CN?. We skip waveforms that are empty, embargoed, or contain over 50 gaps per component.

Once data is acquired, phase arrivals are identified using PhaseNet \citep{zhu2019phasenet}, a deep learning-based algorithm for automatic P- and S-wave detection, trained on the INSTANCE dataset \citep{michelini2021instance}, implemented through the SeisBench framework \citep{munchmeyer2022picker, woollam2022seisbench}. We extract ground velocity estimates around each peak if horizontal components exist. Each job's output includes phase picks, job metadata, and checkpoint information during processing. These are stored in an AWS DocumentDB server, which functions as a NoSQL database for tracking job progress and storing results. This cloud-native architecture supports high throughput, fault-tolerant processing of large-scale seismic datasets, leveraging the scalability and modularity of AWS cloud services — separating data storage (S3), compute (Batch), and output management (DocumentDB).

\subsection{Database}
We employ the AWS DocumentDB to store intermediate and final data products. The DocumentDB service is a NoSQL database that manages and stores data in a structured, flexible format. The database is organized into several collections, each representing a different category of data relevant to the seismic data processing workflows we've implemented here. Each collection contains documents (JSON-like records) with a defined set of fields representing specific information pieces. The collections are also indexed to improve performance and to avoid duplicate picks.  
 
The DocumentDB server organizes station metadata and processing outputs into distinct collections (Table~\ref{tab:docdb}). This design enables efficient querying and auditing of picking results and the exact software and settings used to generate them. The schema is flexible enough to query independent information, as station information, job configuration, and analysis results are stored independently but connected through unique IDs and timestamps. During this experiment, an I/O-optimized class instance (db.r6g.2xlarge, 8 vCPUs, 64 GB RAM) was employed as the DocumentDB server. 

\begin{longtable}{llp{4cm}ll}
\caption{DocumentDB Collections and Schema Summary} \\
\toprule
\textbf{Collection} & \textbf{Field} & \textbf{Description} & \textbf{Type} & \textbf{Example} \\
\midrule
\endfirsthead

\multicolumn{5}{l}{\textit{(continued from previous page)}} \\
\toprule
\textbf{Collection} & \textbf{Field} & \textbf{Description} & \textbf{Type} & \textbf{Example} \\
\midrule
\endhead

\midrule \multicolumn{5}{r}{\textit{(continued on next page)}} \\
\endfoot

\bottomrule
\endlastfoot

\multirow{8}{*}{\texttt{stations}} 
  & \texttt{trace\_id} & Station identifier & str & UW.SHW.01 \\
  & \texttt{network\_code} & Network code & str & UW \\
  & \texttt{station\_code} & Station code & str & SHW \\
  & \texttt{location\_code} & Location code (if any) & str & 01 \\
  & \texttt{channels} & Available channels & str & SH \\
  & \texttt{latitude} & Latitude of station & float & 46.19364 \\
  & \texttt{longitude} & Longitude of station & float & -122.23492 \\
  & \texttt{elevation} & Elevation in meters & float & 1442.0 \\
  & \texttt{start\_date} & Operational start date & str & 1972.275 \\
  & \texttt{end\_date} & Operational end date & str & 3000.001 \\

\midrule

\multirow{7}{*}{\texttt{picks}} 
  & \texttt{trace\_id} & Station identifier & str & NC.NFR. \\
  & \texttt{start\_time} & Start of pick window & datetime & 2012-01-01T01:08:35.800 \\
  & \texttt{peak\_time} & Peak amplitude time & datetime & 2012-01-01T01:08:36.000 \\
  & \texttt{end\_time} & End of pick window & datetime & 2012-01-01T01:08:36.200 \\
  & \texttt{confidence} & ML confidence score & float & 0.2743 \\
  & \texttt{amplitude} & Signal amplitude & float & 0.000013 \\
  & \texttt{phase} & Phase type (P/S) & str & S \\
  & \texttt{run\_id} & Associated run ID & ObjectId & ObjectId(...) \\

\midrule

\multirow{6}{*}{\texttt{picks\_record}} 
  & \texttt{trace\_id} & Station identifier & str & NC.NFR. \\
  & \texttt{channel} & Channel code & str & HH \\
  & \texttt{year} & Year of record & int & 2012 \\
  & \texttt{doy} & Day of year & int & 1 \\
  & \texttt{n\_picks} & Number of picks & int & 573 \\
  & \texttt{run\_id} & Associated run ID & ObjectId & ObjectId(...) \\

\midrule

\multirow{6}{*}{\texttt{sb\_runs}} 
  & \texttt{run\_id} & Run ID & ObjectId & ObjectId(...) \\
  & \texttt{model} & ML model used & str & PhaseNet \\
  & \texttt{weight} & Model weight name & str & instance \\
  & \texttt{p\_threshold} & P-phase detection threshold & float & 0.2 \\
  & \texttt{s\_threshold} & S-phase detection threshold & float & 0.2 \\
  & \texttt{components\_load} & Input component configure & str & ZNE12 \\
  & \texttt{seisbench\_ver} & SeisBench version & str & 0.8.2 \\
  & \texttt{weight\_ver} & Weight version & str & 2 \\
\label{tab:docdb}
\end{longtable}

\section{Results}

\subsection{Job Statistics}
\begin{figure}[ht!]
    \includegraphics[width=\textwidth]{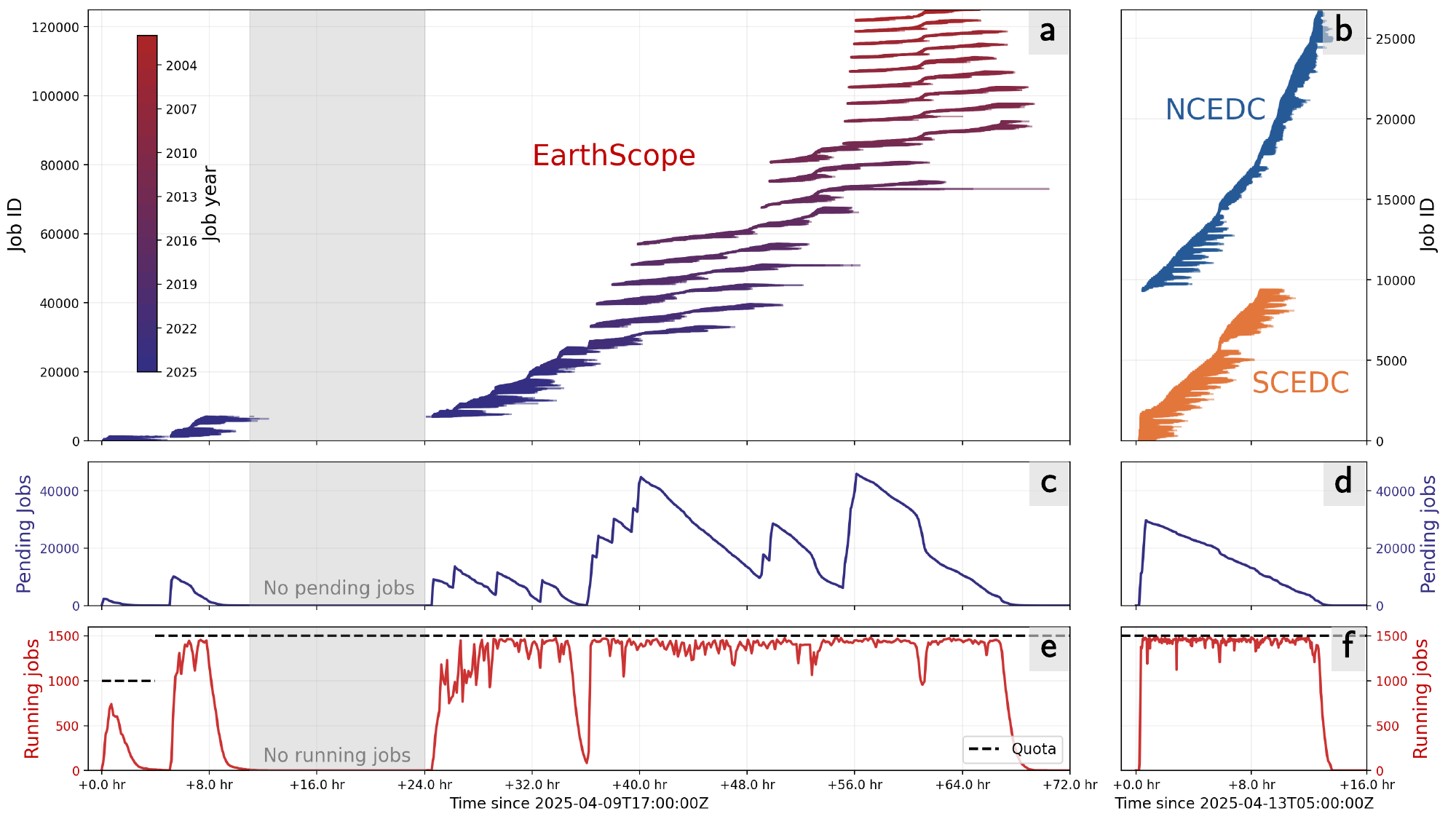} 
    \caption{Detailed job run history for the EarthScope, NCEDC, and SCEDC dataset. Panel (a) shows the Job ID as a function of time, color-coded by year the data was recorded, for the EarthScope dataset. Panel (b) shows the job progression for the NCEDC and SCEDC datasets. Panels (c) and (d) show the pending jobs as a function of time. Panels (e) and (f) show the running jobs as a function of time. The horizontal dashed line represents the job quota.}
    \label{fig:3}
\end{figure}

Mining the EarthScope dataset took less than three days, while the combined NCEDC and SCEDC datasets took less than 16 hours. Since this was the first major data mining exercise on the EarthScope archive, job sets were launched manually, progressively, and actively monitored. Each set of jobs was intended to process a year of data recorded on all stations with channels matching those listed above. Figure~\ref{fig:3}a shows the progression of the jobs mining the EarthScope dataset, color-coded by the year the data was recorded. Figure~\ref{fig:3}c and e show the pending and running jobs as a function of time. Time periods where jobs were manually launched manifest as increases in the number of pending and running jobs. Since we used a quota of 1500 running jobs, the pending jobs decreased at an approximately constant rate. There are occasional abrupt decreases in the number of running jobs, for example, the one near 60 hours in Figure~\ref{fig:3}c, that correspond to times when stations are listed as available in the metadata service, but no data is actually loaded. Figure~\ref{fig:3}b, d, and f show the job ID, pending, and running as a function of time. For the NCEDC and SCEDC datasets, all jobs were submitted and queued simultaneously and not parsed by year, as they were with the EarthScope dataset.

\subsection{Phase Picking}
The mining exercise resulted in a total of 4.3 billion picks (2.8 billion P-wave picks, 1.5 billion S-wave picks), with the EarthScope, NCEDC, and SCEDC datasets containing 2.8, 1.1, and 0.4 billion picks, respectively. Figure S1 shows the total number of picks broken down by the dominant network codes in each dataset. The NC network has the largest number of picks at nearly 1 billion, followed by UW at 0.5 billion, and CI at 0.4 billion.

\begin{figure}[ht!]
    \includegraphics[width=\textwidth]{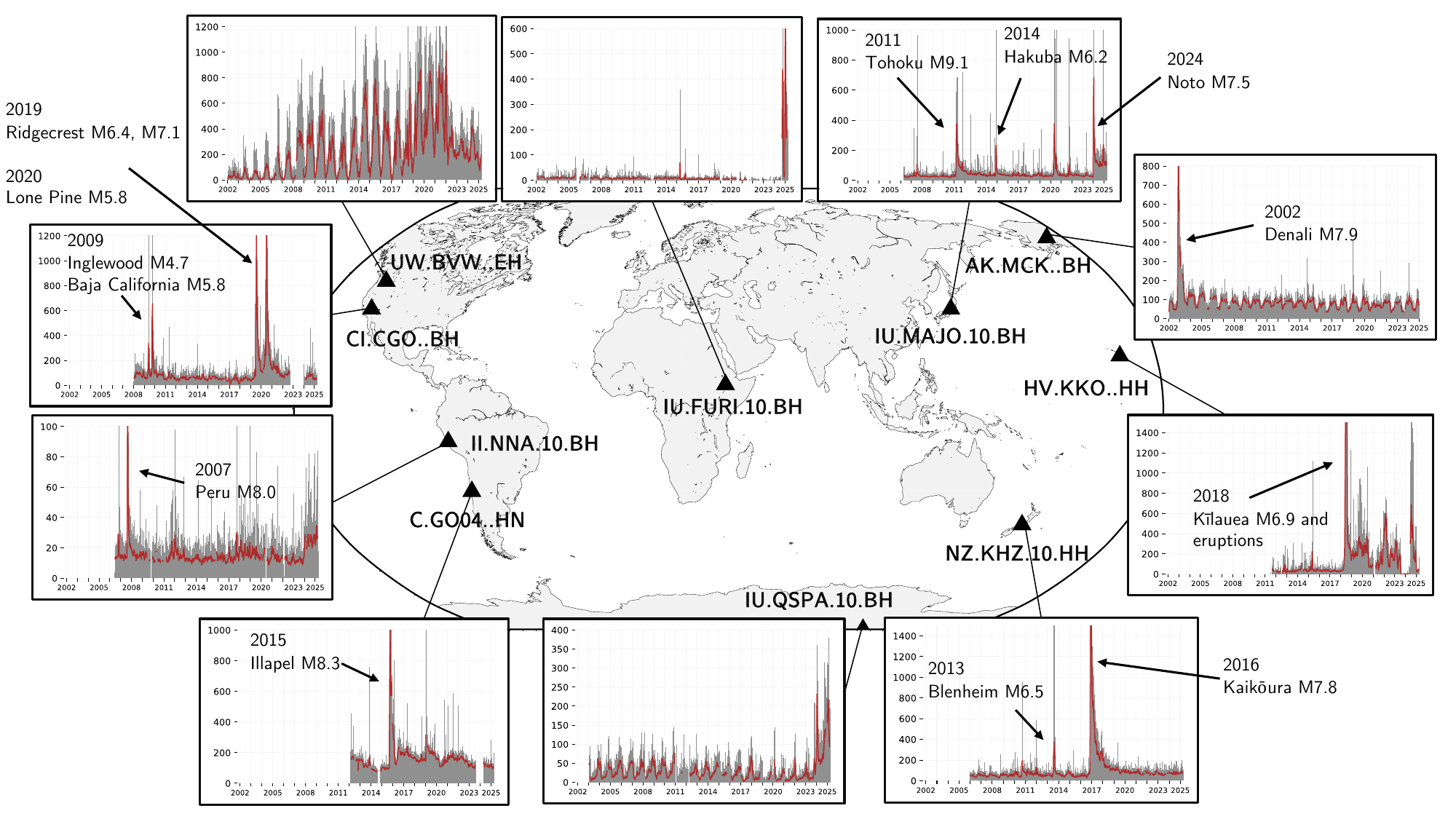} 
    \caption{Daily picks for selected stations. Stations are indicated by triangles on the central map, annotated with location and channel codes. For each of the ten example stations, the detail plots show a time series of the number of picks per day and a 28-day moving average in red.}
    \label{fig:4}
\end{figure}

Figure~\ref{fig:4} shows examples of the daily number of picks as a function of time for ten stations around the world. Station UW.BVW is a station near Beverly, WA, atop the Saddle Mountains. UW.BVW displays a strong seasonality of detections, with detection rates peaking in the late summer. IU.FURI is a station in southern Addis Ababa, Ethiopia. This station shows a large number of detections beginning in late 2024 and extending into early 2025. These detections likely correspond to a swarm of earthquakes that began in late September and produced 19 M5+ earthquakes, the largest of which was a M5.9 on February 14th, 2025. IU.MAJO in Matsushiro, Japan, clearly recorded aftershocks from the 2011 M9 Tohoku-oki, 2014 M6.2 Hakuba, and 2024 M7.5 Noto earthquakes as well as many others. AK.MCK in McKinley Park, AK shows strong seasonality in detections and recorded the 2002 M7.9 Denali earthquake. HV.KKO in Hawai'i records seismicity from the 2018 M6.9 earthquake and ongoing eruptive activity from the Kilauea volcano. NZ.KHZ in Kahutara, New Zealand, recorded the 2013 M6.5 Blenheim and the 2016 M7.8 Kaikōura earthquakes and aftershocks. IU.QSPA at the South Pole shows a strong seasonality of detections and also records increased detections beginning in late 2024, possibly due to a series of M5+ events near the Balleny Islands and on the Pacific-Antarctic Ridge. C.GO04 at the Tololo Observatory, Vicuna, Chile, records the 2015 M8.3 Illapel and the M6.7 Coquimbo earthquakes. II.NNA records the 2007 M8.0 Pisco, Peru earthquake. Finally, CI.CGO records the 2009 M4.7 Inglewood, 2009 M5.8 Northern Baja California, the M6.4 and M7.1 2019 Ridgecrest, and the 2020 M5.8 Lone Pine earthquakes.

\begin{figure}[ht!]
    \includegraphics[width=\textwidth]{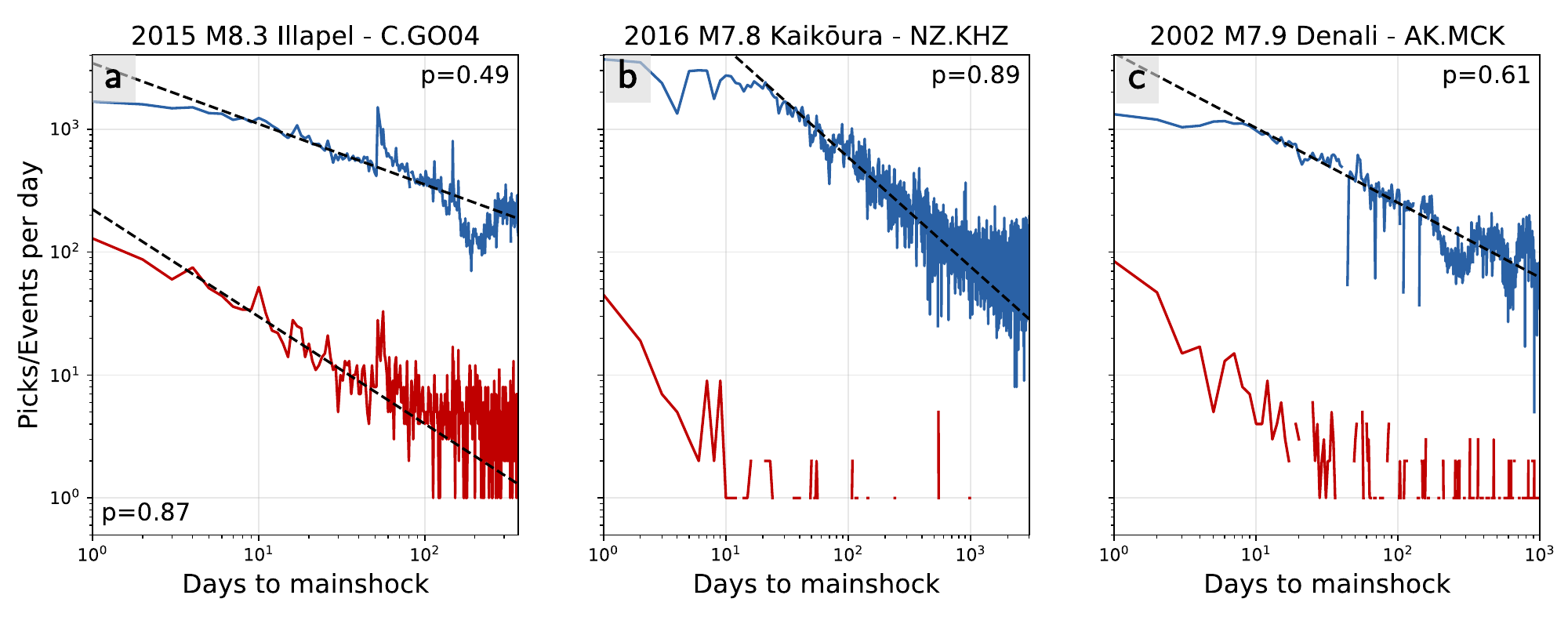} 
    \caption{Omori-type decays of the number of picks per day (blue) and the number of events per day (red). We use picks at the reference station provided in the figure title. Event counts for the Illapel earthquake are from the Chilean Seismic Network (CSN) catalog, for the other two examples from the International Seismological Centre (ISC) and United States Geological Survey (USGS) catalog. In each case, we count all events with at most a 1.5-degree difference in latitude and longitude from the reference station.}
    \label{fig:omori}
\end{figure}

Figure~\ref{fig:omori} shows the number of picks and earthquakes after three selected large earthquakes: (a) 2015 M8.3 Illapel earthquake, (b) 2016 M7.8 Kaikōura earthquake, and (c) 2002 M7.9 Denali earthquake. In each case, the daily pick count follows approximately an Omori decay law \citep{utsu1961statistical}. While the Omori decay in the event count can only be observed for a short period in the regional and global reference catalogs used, they are stable over longer durations in the pick count. For the Kaikōura earthquake, the Omori decay in pick counts is stable over more than 1000 days. Notably, the Omori $p$ values, describing the decay rate, differ between picks and events with systematically higher $p$ values for events, indicating a faster decay. For example, for the Illapel earthquake, we estimate $p=0.87$ for the event counts, a typical value, yet $p=0.49$ for the pick counts, a surprisingly low value. We suggest this difference originates from the joint effects of event rate and event magnitude on the pick counts. This highlights the value of analyzing pick count dynamics to study earthquake statistics. Several time series of picks shown in Figure~\ref{fig:4} exhibit seasonality, demonstrating the potential effects of the noise floor. 

\section{Discussion}

To estimate the total number of earthquakes we can identify from the 4.3 billion seismic phase picks, we begin by assessing how picks are distributed across different magnitude bins. Based on estimates from \citet{mcbrearty2023earthquake}, we expect an average of approximately 20 P-wave picks per M1 earthquake, 60 picks per M2, and 100 picks per M3. Because earthquake frequency decreases with increasing magnitude, for each M1 event (20 picks), we expect roughly 0.1 M2 events (6 picks) and 0.01 M3 events (1 pick). This implies that M2s contribute only about 30\% as many picks as M1s, and M3s contribute just 5\%. These approximate ratios are consistent with those observed in other deep learning-based catalogs (e.g., \citet{munchmeyer2025characterising}), suggesting that the dataset is overwhelmingly composed of picks associated with $\sim$M1 earthquakes and smaller.

Using this assumption, and taking 20 picks per M1 event as a working average, along with an estimated conservative association rate of 25\%, we arrive at a rough estimate of more than 54 million earthquakes that can be associated from the data. We note that many further picks will correspond to actual events, that just lack sufficiently many detections to successfully associate them. For context, the statewide California catalog contains approximately 325,000 events \citep{zhu2025california}, while the Advanced National Seismic System (ANSS) Comprehensive Earthquake Catalog (ComCat) includes about 4 million events for the United States and larger global events. Deep learning methods have already demonstrated the ability to increase catalog completeness by up to an order of magnitude in sparsely instrumented regions \citep{park2022basement}. Given the unprecedented scale of this pick database, we fully expect it to significantly expand the number of earthquakes identified globally, particularly in regions that have historically been under-detected.

We anticipate that the archive of picks will be of use to many researchers in the field of earthquake science. As such, we have made it available for public query using web services and URL builders (see \hyperref[sec:availability]{Data and Code Availability}). A dedicated EC2 instance receives HTTPS requests and directly returns query results. As an example, one can query the picking results for AK.MCK. station for one month of data using the following Python script. The script uses the pandas package (\url{https://pandas.pydata.org}) and loads the query results as a data frame. We also offer a binary dump of the entire database for download.

\begin{lstlisting}[caption=Python query of the picking database, label=code, language=Python]
import pandas as pd
base_url = "https://dasway.ess.washington.edu/quakescope/service/picks/"
url = f"{base_url}query?tid=AK.MCK.&start_time=2010-01-01&end_time=2010-02-01&limit=1000"
pick = pd.read_csv(url, delimiter="|")
\end{lstlisting}

Our demonstration is simply a starting point for future and more focused exploration. For instance, we did omit the NCEDC SP? channels that had a name change at some time during the network operation. We also omitted using specialized base models, such as PickBlue that include hydrophone data and would be more appropriate for picking ocean-bottom seismometer (OBS) data \citep{bornstein2024pickblue}. There are also opportunities to identify non-traditional seismicity, such as volcanic earthquakes \citep{zhong2024volcano} or low-frequency earthquakes associated with slow slip events \citep{munchmeyer2023deep, lin2024detection}. The fully open-source and modular design of the workflow ensures reproducibility while allowing flexibility to incorporate different models and pre-trained weights optimized for different applications. This enables other researchers to easily adopt and extend the cloud-native workflow for their own analyses. Future work might consider using template matching to enhance our detections, as done in California \citep{ross2019searching}.

Moreover, we did not use data from other seismic data providers. For instance, we omitted the non-FDSN seismic networks, the Observatories and Research Facilities for European Seismology (ORFEUS) federated networks, and the AusPass networks. Pulling data from these data centers can be done using their FDSN web services; however, care must be taken not to overload the data service \citep[e.g.,][]{maccarthy2020cloud}. Because of the resiliency of the commercial cloud providers to spiked demand in data access, a strategy to improve the stability of the non-cloud-hosted archive is to use cloud storage as a backup in case of service interruption and route users toward the cloud-hosted archive instead of the local seismic network's data servers.

\section{Conclusions}

In conclusion, our data mining experiment plants the seed for impactful advances in geophysics. Cloud-based and AI-aided picking of P- and S-waves can be used to retrain neural networks and improve the rapid and precise assessment of earthquake hazards, such as earthquake early warning \citep[e.g.,][]{zhang2024universal}. These newly detected potential earthquakes and waveforms may be valuable to build foundational models for seismology that learn fundamental seismic signal patterns from massive waveform libraries \citep[e.g.,][]{wang2025seismollm, liu2024seislm}, and then be fine-tuned for specific tasks (e.g., picking, polarity, backazimuth, etc.). Running petabyte-scale workflow on the cloud also provides a testbed for greener computing with tools that can allow researchers to deploy carbon-aware computing jobs \citep[e.g.,][]{west2025exploring}

\begin{acknowledgements}
This work is supported by the Seismic Computational Platform for Empowering Discovery (SCOPED) project under the National Science Foundation (award numbers OAC-2103701 (UW), OAC-2103494 (UT)). YN and MD are also partially supported by the EarthScope Consortium through a Pass-Through Entity (PTE) Federal award no 2310069. The computing resources presented in this paper were obtained using CloudBank \citep{norman2021cloudbank}, which is supported by the National Science Foundation (award number CNS-1925001). Data were accessed from the NSF SAGE data repository operated by EarthScope Consortium (award number 1724509). JM has been funded by the European Union under the grant agreement n°101104996 (DECODE). The Cascadia Region Earthquake Science Center (CRESCENT), funded by National Science Foundation Cooperative Agreement \#2225286, partially supported this project through funding to AMT and LB. AMT was also supported by the National Science Foundation (award number 1848302).
\end{acknowledgements}
		
\section*{Data and Code Availability}\label{sec:availability}
Data from SCEDC and NCEDC is publicly available on AWS S3, while data hosted at EarthScope could be accessed through the EarthScope FDSN web service. The DOIs of FDSN seismic networks used in this study are listed at \url{https://dasway.ess.washington.edu/quakescope/doi.csv} and added as supplementary materials. All data products are hosted at \url{https://dasway.ess.washington.edu/quakescope}. URL builders are available for database queries. The database snapshots are also dumped as BSON (binary JSON) files. Our code base is hosted on GitHub at \url{https://github.com/SeisSCOPED/QuakeScope}.
		
\section*{Competing Interests}
The authors declare no competing interests.

\bibliography{mybibfile}

\begin{thebibliography}{42}
\providecommand{\natexlab}[1]{#1}
\providecommand{\url}[1]{\texttt{#1}}
\expandafter\ifx\csname urlstyle\endcsname\relax
  \providecommand{\doi}[1]{doi: #1}\else
  \providecommand{\doi}{doi: \begingroup \urlstyle{rm}\Url}\fi

\bibitem[Allen(1982)]{allen1982automatic}
Allen, R.
\newblock Automatic phase pickers: Their present use and future prospects.
\newblock \emph{Bulletin of the Seismological Society of America}, 72\penalty0 (6B):\penalty0 S225--S242, 12 1982.
\newblock \doi{https://doi.org/10.1785/BSSA07206B0225}.

\bibitem[Beyreuther et~al.(2010)Beyreuther, Barsch, Krischer, Megies, Behr, and Wassermann]{beyreuther2010obspy}
Beyreuther, M., Barsch, R., Krischer, L., Megies, T., Behr, Y., and Wassermann, J.
\newblock ObsPy: A Python toolbox for seismology.
\newblock \emph{Seismological Research Letters}, 81\penalty0 (3):\penalty0 530--533, 2010.
\newblock \doi{https://doi.org/10.1785/gssrl.81.3.530}.

\bibitem[Bornstein et~al.(2024)Bornstein, Lange, M{\"u}nchmeyer, Woollam, Rietbrock, Barcheck, Grevemeyer, and Tilmann]{bornstein2024pickblue}
Bornstein, T., Lange, D., M{\"u}nchmeyer, J., Woollam, J., Rietbrock, A., Barcheck, G., Grevemeyer, I., and Tilmann, F.
\newblock PickBlue: Seismic phase picking for ocean bottom seismometers with deep learning.
\newblock \emph{Earth and Space Science}, 11\penalty0 (1):\penalty0 e2023EA003332, 2024.
\newblock \doi{https://doi.org/10.1029/2023EA003332}.

\bibitem[Gentemann et~al.(2021)Gentemann, Holdgraf, Abernathey, Crichton, Colliander, Kearns, Panda, and Signell]{gentemann2021science}
Gentemann, C.~L., Holdgraf, C., Abernathey, R., Crichton, D., Colliander, J., Kearns, E.~J., Panda, Y., and Signell, R.~P.
\newblock Science storms the cloud.
\newblock \emph{AGU Advances}, 2\penalty0 (2):\penalty0 e2020AV000354, 2021.
\newblock \doi{https://doi.org/10.1029/2020AV000354}.

\bibitem[Hibert et~al.(2014)Hibert, Mangeney, Grandjean, Baillard, Rivet, Shapiro, Satriano, Maggi, Boissier, Ferrazzini, et~al.]{hibert2014automated}
Hibert, C., Mangeney, A., Grandjean, G., Baillard, C., Rivet, D., Shapiro, N.~M., Satriano, C., Maggi, A., Boissier, P., Ferrazzini, V., et~al.
\newblock Automated identification, location, and volume estimation of rockfalls at Piton de la Fournaise volcano.
\newblock \emph{Journal of Geophysical Research: Earth Surface}, 119\penalty0 (5):\penalty0 1082--1105, 2014.
\newblock \doi{https://doi.org/10.1002/2013JF002970}.

\bibitem[Krauss et~al.(2023)Krauss, Ni, Henderson, and Denolle]{krauss2023cloud}
Krauss, Z., Ni, Y., Henderson, S., and Denolle, M.
\newblock Seismology in the cloud: guidance for the individual researcher.
\newblock \emph{Seismica}, 2\penalty0 (2), 08 2023.
\newblock \doi{https://doi.org/10.26443/seismica.v2i2.979}.

\bibitem[Lin et~al.(2024)Lin, Thomas, Bachelot, Toomey, Searcy, and Melgar]{lin2024detection}
Lin, J.-T., Thomas, A.~M., Bachelot, L., Toomey, D., Searcy, J., and Melgar, D.
\newblock Detection of Hidden Low-Frequency Earthquakes in Southern Vancouver Island with Deep Learning.
\newblock \emph{Seismica}, 2\penalty0 (4), 10 2024.
\newblock \doi{https://doi.org/10.26443/seismica.v2i4.1134}.

\bibitem[Liu et~al.(2024)Liu, M{\"u}nchmeyer, Laurenti, Marone, de~Hoop, and Dokmani{\'c}]{liu2024seislm}
Liu, T., M{\"u}nchmeyer, J., Laurenti, L., Marone, C., de~Hoop, M.~V., and Dokmani{\'c}, I.
\newblock SeisLM: a Foundation Model for Seismic Waveforms.
\newblock \emph{arXiv preprint arXiv:2410.15765}, 2024.
\newblock \doi{https://doi.org/10.48550/arXiv.2410.15765}.

\bibitem[MacCarthy et~al.(2020)MacCarthy, Marcillo, and Trabant]{maccarthy2020cloud}
MacCarthy, J., Marcillo, O., and Trabant, C.
\newblock Seismology in the Cloud: A New Streaming Workflow.
\newblock \emph{Seismological Research Letters}, 91\penalty0 (3):\penalty0 1804--1812, 03 2020.
\newblock \doi{https://doi.org/10.1785/0220190357}.

\bibitem[McBrearty and Beroza(2023)]{mcbrearty2023earthquake}
McBrearty, I.~W. and Beroza, G.~C.
\newblock Earthquake phase association with graph neural networks.
\newblock \emph{Bulletin of the Seismological Society of America}, 113\penalty0 (2):\penalty0 524--547, 2023.
\newblock \doi{https://doi.org/10.1785/0120220182}.

\bibitem[Michelini et~al.(2021)Michelini, Cianetti, Gaviano, Giunchi, Jozinovi{\'c}, and Lauciani]{michelini2021instance}
Michelini, A., Cianetti, S., Gaviano, S., Giunchi, C., Jozinovi{\'c}, D., and Lauciani, V.
\newblock INSTANCE--the Italian seismic dataset for machine learning.
\newblock \emph{Earth System Science Data}, 13\penalty0 (12):\penalty0 5509--5544, 2021.
\newblock \doi{https://doi.org/10.5194/essd-13-5509-2021}.

\bibitem[Mousavi et~al.(2019{\natexlab{a}})Mousavi, Sheng, Zhu, and Beroza]{mousavi2019stanford}
Mousavi, S.~M., Sheng, Y., Zhu, W., and Beroza, G.~C.
\newblock STanford EArthquake Dataset (STEAD): A global data set of seismic signals for AI.
\newblock \emph{IEEE Access}, 7:\penalty0 179464--179476, 2019{\natexlab{a}}.
\newblock \doi{https://doi.org/10.1109/ACCESS.2019.2947848}.

\bibitem[Mousavi et~al.(2019{\natexlab{b}})Mousavi, Zhu, Sheng, and Beroza]{mousavi2019cred}
Mousavi, S.~M., Zhu, W., Sheng, Y., and Beroza, G.~C.
\newblock CRED: A deep residual network of convolutional and recurrent units for earthquake signal detection.
\newblock \emph{Scientific reports}, 9\penalty0 (1):\penalty0 10267, 2019{\natexlab{b}}.
\newblock \doi{https://doi.org/10.1038/s41598-019-45748-1}.

\bibitem[Mousavi et~al.(2020)Mousavi, Ellsworth, Zhu, Chuang, and Beroza]{mousavi2020earthquake}
Mousavi, S.~M., Ellsworth, W.~L., Zhu, W., Chuang, L.~Y., and Beroza, G.~C.
\newblock Earthquake transformer—an attentive deep-learning model for simultaneous earthquake detection and phase picking.
\newblock \emph{Nature Communications}, 11\penalty0 (1):\penalty0 3952, 2020.
\newblock \doi{https://doi.org/10.1038/s41467-020-17591-w}.

\bibitem[M{\"u}nchmeyer et~al.(2022)M{\"u}nchmeyer, Woollam, Rietbrock, Tilmann, Lange, Bornstein, Diehl, Giunchi, Haslinger, Jozinovi{\'c}, et~al.]{munchmeyer2022picker}
M{\"u}nchmeyer, J., Woollam, J., Rietbrock, A., Tilmann, F., Lange, D., Bornstein, T., Diehl, T., Giunchi, C., Haslinger, F., Jozinovi{\'c}, D., et~al.
\newblock Which picker fits my data? A quantitative evaluation of deep learning based seismic pickers.
\newblock \emph{Journal of Geophysical Research: Solid Earth}, 127\penalty0 (1):\penalty0 e2021JB023499, 2022.
\newblock \doi{https://doi.org/10.1029/2021JB023499}.

\bibitem[M{\"u}nchmeyer et~al.(2025)M{\"u}nchmeyer, Molina-Ormazabal, Marsan, Langlais, Baez, Heit, Gonz{\'a}lez-Vidal, Moreno, Tilmann, Lange, et~al.]{munchmeyer2025characterising}
M{\"u}nchmeyer, J., Molina-Ormazabal, D., Marsan, D., Langlais, M., Baez, J.-C., Heit, B., Gonz{\'a}lez-Vidal, D., Moreno, M., Tilmann, F., Lange, D., et~al.
\newblock Characterising the Atacama segment of the Chile subduction margin (24 S-31 S) with> 165,000 earthquakes.
\newblock \emph{arXiv preprint arXiv:2501.14396}, 2025.
\newblock \doi{https://doi.org/10.48550/arXiv.2501.14396}.

\bibitem[Münchmeyer et~al.(2024)Münchmeyer, Giffard-Roisin, Malfante, Frank, Poli, Marsan, and Socquet]{munchmeyer2023deep}
Münchmeyer, J., Giffard-Roisin, S., Malfante, M., Frank, W., Poli, P., Marsan, D., and Socquet, A.
\newblock Deep learning detects uncataloged low-frequency earthquakes across regions.
\newblock \emph{Seismica}, 3\penalty0 (1), 05 2024.
\newblock \doi{https://doi.org/10.26443/seismica.v3i1.1185}.

\bibitem[Ni et~al.(2023)Ni, Hutko, Skene, Denolle, Malone, Bodin, Hartog, and Wright]{ni2023curated}
Ni, Y., Hutko, A., Skene, F., Denolle, M., Malone, S., Bodin, P., Hartog, R., and Wright, A.
\newblock Curated Pacific Northwest AI-ready Seismic Dataset.
\newblock \emph{Seismica}, 2\penalty0 (1), May 2023.
\newblock \doi{https://doi.org/10.26443/seismica.v2i1.368}.

\bibitem[Norman et~al.(2021)Norman, Kellen, Smallen, DeMeulle, Strande, Lazowska, Alterman, Fatland, Stone, Tan, Yelick, Van~Dusen, and Mitchell]{norman2021cloudbank}
Norman, M., Kellen, V., Smallen, S., DeMeulle, B., Strande, S., Lazowska, E., Alterman, N., Fatland, R., Stone, S., Tan, A., Yelick, K., Van~Dusen, E., and Mitchell, J.
\newblock CloudBank: Managed Services to Simplify Cloud Access for Computer Science Research and Education.
\newblock In \emph{Practice and Experience in Advanced Research Computing 2021: Evolution Across All Dimensions}, PEARC '21. Association for Computing Machinery, 2021.
\newblock \doi{https://doi.org/10.1145/3437359.3465586}.

\bibitem[Park et~al.(2022)Park, Beroza, and Ellsworth]{park2022basement}
Park, Y., Beroza, G.~C., and Ellsworth, W.~L.
\newblock Basement Fault Activation before Larger Earthquakes in Oklahoma and Kansas.
\newblock \emph{The Seismic Record}, 2\penalty0 (3):\penalty0 197--206, 08 2022.
\newblock \doi{https://doi.org/10.1785/0320220020}.

\bibitem[Perol et~al.(2018)Perol, Gharbi, and Denolle]{perol2018convolutional}
Perol, T., Gharbi, M.~J., and Denolle, M.
\newblock Convolutional neural network for earthquake detection and location.
\newblock \emph{Science Advances}, 4\penalty0 (2):\penalty0 e1700578, 2018.
\newblock \doi{https://doi.org/10.1126/sciadv.1700578}.

\bibitem[Retailleau et~al.(2022)Retailleau, Saurel, Zhu, Satriano, Beroza, Issartel, Boissier, Team, Team, et~al.]{retailleau2022wrapper}
Retailleau, L., Saurel, J.-M., Zhu, W., Satriano, C., Beroza, G.~C., Issartel, S., Boissier, P., Team, O., Team, O., et~al.
\newblock A wrapper to use a machine-learning-based algorithm for earthquake monitoring.
\newblock \emph{Seismological Research Letters}, 93\penalty0 (3):\penalty0 1673--1682, 2022.
\newblock \doi{https://doi.org/10.1785/0220210279}.

\bibitem[Ross et~al.(2018)Ross, Meier, Hauksson, and Heaton]{ross2018generalized}
Ross, Z.~E., Meier, M.-A., Hauksson, E., and Heaton, T.~H.
\newblock Generalized seismic phase detection with deep learning.
\newblock \emph{Bulletin of the Seismological Society of America}, 108\penalty0 (5A):\penalty0 2894--2901, 2018.
\newblock \doi{https://doi.org/10.1785/0120180080}.

\bibitem[Ross et~al.(2019{\natexlab{a}})Ross, Trugman, Hauksson, and Shearer]{ross2019searching}
Ross, Z.~E., Trugman, D.~T., Hauksson, E., and Shearer, P.~M.
\newblock Searching for hidden earthquakes in Southern California.
\newblock \emph{Science}, 364\penalty0 (6442):\penalty0 767--771, 2019{\natexlab{a}}.
\newblock \doi{https://doi.org/10.1126/science.aaw6888}.

\bibitem[Ross et~al.(2019{\natexlab{b}})Ross, Yue, Meier, Hauksson, and Heaton]{ross2019phaselink}
Ross, Z.~E., Yue, Y., Meier, M.-A., Hauksson, E., and Heaton, T.~H.
\newblock PhaseLink: A deep learning approach to seismic phase association.
\newblock \emph{Journal of Geophysical Research: Solid Earth}, 124\penalty0 (1):\penalty0 856--869, 2019{\natexlab{b}}.
\newblock \doi{https://doi.org/10.1029/2018JB016674}.

\bibitem[Ross et~al.(2020)Ross, Meier, Hauksson, and Heaton]{ross2020p}
Ross, Z.~E., Meier, M.-A., Hauksson, E., and Heaton, T.~H.
\newblock P-wave arrival picking and first-motion polarity determination with deep learning.
\newblock \emph{Journal of Geophysical Research: Solid Earth}, 125\penalty0 (4):\penalty0 e2019JB018663, 2020.
\newblock \doi{https://doi.org/10.1029/2017JB015251}.

\bibitem[Sun et~al.(2024)Sun, Pan, Huang, Guan, Yen, Ho, Chi, Ku, Huang, Fu, et~al.]{sun2024deep}
Sun, W.-F., Pan, S.-Y., Huang, C.-M., Guan, Z.-K., Yen, I.-C., Ho, C.-W., Chi, T.-C., Ku, C.-S., Huang, B.-S., Fu, C.-C., et~al.
\newblock Deep learning-based earthquake catalog reveals the seismogenic structures of the 2022 MW 6.9 Chihshang earthquake sequence.
\newblock \emph{Terrestrial, Atmospheric and Oceanic Sciences}, 35\penalty0 (1):\penalty0 5, 2024.
\newblock \doi{https://doi.org/10.1007/s44195-024-00063-9}.

\bibitem[Utsu(1961)]{utsu1961statistical}
Utsu, T.
\newblock A statistical study on the occurrence of aftershocks.
\newblock \emph{Geophys. Mag.}, 30:\penalty0 521--605, 1961.

\bibitem[Walter et~al.(2021)Walter, Ogwari, Thiel, Ferrer, and Woelfel]{walter2021easyquake}
Walter, J.~I., Ogwari, P., Thiel, A., Ferrer, F., and Woelfel, I.
\newblock easyQuake: Putting machine learning to work for your regional seismic network or local earthquake study.
\newblock \emph{Seismological Society of America}, 92\penalty0 (1):\penalty0 555--563, 2021.
\newblock \doi{https://doi.org/10.1785/0220200226}.

\bibitem[Wang et~al.(2025)Wang, Liu, Su, Wang, Bai, and Ouyang]{wang2025seismollm}
Wang, X., Liu, F., Su, R., Wang, Z., Bai, L., and Ouyang, W.
\newblock SeisMoLLM: Advancing Seismic Monitoring via Cross-modal Transfer with Pre-trained Large Language Model.
\newblock \emph{arXiv preprint arXiv:2502.19960}, 2025.
\newblock \doi{https://doi.org/10.48550/arXiv.2502.19960}.

\bibitem[West et~al.(2025)West, Lehmann, Bountris, Leser, Elkhatib, and Thamsen]{west2025exploring}
West, K., Lehmann, F., Bountris, V., Leser, U., Elkhatib, Y., and Thamsen, L.
\newblock Exploring the Potential of Carbon-Aware Execution for Scientific Workflows.
\newblock \emph{arXiv preprint arXiv:2503.13705}, 2025.
\newblock \doi{https://doi.org/10.48550/arXiv.2503.13705}.

\bibitem[Woollam et~al.(2022)Woollam, M{\"u}nchmeyer, Tilmann, Rietbrock, Lange, Bornstein, Diehl, Giunchi, Haslinger, Jozinovi{\'c}, et~al.]{woollam2022seisbench}
Woollam, J., M{\"u}nchmeyer, J., Tilmann, F., Rietbrock, A., Lange, D., Bornstein, T., Diehl, T., Giunchi, C., Haslinger, F., Jozinovi{\'c}, D., et~al.
\newblock SeisBench—A toolbox for machine learning in seismology.
\newblock \emph{Seismological Society of America}, 93\penalty0 (3):\penalty0 1695--1709, 2022.
\newblock \doi{https://doi.org/10.1785/0220210324}.

\bibitem[Yeck et~al.(2021)Yeck, Patton, Ross, Hayes, Guy, Ambruz, Shelly, Benz, and Earle]{yeck2021leveraging}
Yeck, W.~L., Patton, J.~M., Ross, Z.~E., Hayes, G.~P., Guy, M.~R., Ambruz, N.~B., Shelly, D.~R., Benz, H.~M., and Earle, P.~S.
\newblock Leveraging deep learning in global 24/7 real-time earthquake monitoring at the National Earthquake Information Center.
\newblock \emph{Seismological Society of America}, 92\penalty0 (1):\penalty0 469--480, 2021.
\newblock \doi{https://doi.org/10.1785/0220200178}.

\bibitem[Yu et~al.(2021)Yu, Bhaskaran, Chen, Ross, Hauksson, and Clayton]{yu2021scedc}
Yu, E., Bhaskaran, A., Chen, S., Ross, Z.~E., Hauksson, E., and Clayton, R.~W.
\newblock Southern California Earthquake Data Now Available in the AWS Cloud.
\newblock \emph{Seismological Research Letters}, 92\penalty0 (5):\penalty0 3238--3247, 06 2021.
\newblock \doi{https://doi.org/10.1785/0220210039}.

\bibitem[Zawacki et~al.(2023)Zawacki, Bendick, and Woodward]{zawacki2023advancing}
Zawacki, E.~E., Bendick, R., and Woodward, R.~L.
\newblock Advancing geophysics: IRIS and UNAVCO merge to form EarthScope Consortium, 2023.
\newblock \doi{https://doi.org/10.1029/2023CN000227}.

\bibitem[Zhang et~al.(2022)Zhang, Liu, Feng, Wang, and Zhu]{zhang2022loc}
Zhang, M., Liu, M., Feng, T., Wang, R., and Zhu, W.
\newblock LOC-FLOW: An end-to-end machine learning-based high-precision earthquake location workflow.
\newblock \emph{Seismological Society of America}, 93\penalty0 (5):\penalty0 2426--2438, 2022.
\newblock \doi{https://doi.org/10.1785/0220220019}.

\bibitem[Zhang and Zhang(2024)]{zhang2024universal}
Zhang, X. and Zhang, M.
\newblock Universal neural networks for real-time earthquake early warning trained with generalized earthquakes.
\newblock \emph{Communications Earth \& Environment}, 5\penalty0 (1):\penalty0 528, 2024.
\newblock \doi{https://doi.org/10.1038/s43247-024-01718-8}.

\bibitem[Zhong and Tan(2024)]{zhong2024volcano}
Zhong, Y. and Tan, Y.~J.
\newblock Deep-Learning-Based Phase Picking for Volcano-Tectonic and Long-Period Earthquakes.
\newblock \emph{Geophysical Research Letters}, 51\penalty0 (12):\penalty0 e2024GL108438, 2024.
\newblock \doi{https://doi.org/10.1029/2024GL108438}.

\bibitem[Zhu and Beroza(2019)]{zhu2019phasenet}
Zhu, W. and Beroza, G.~C.
\newblock Phasenet: a deep-neural-network-based seismic arrival time picking method.
\newblock \emph{Geophysical Journal International}, 216\penalty0 (1):\penalty0 261--273, 2019.
\newblock \doi{https://doi.org/10.1093/gji/ggy423}.

\bibitem[Zhu et~al.(2022)Zhu, McBrearty, Mousavi, Ellsworth, and Beroza]{zhu2022earthquake}
Zhu, W., McBrearty, I.~W., Mousavi, S.~M., Ellsworth, W.~L., and Beroza, G.~C.
\newblock Earthquake phase association using a Bayesian Gaussian mixture model.
\newblock \emph{Journal of Geophysical Research: Solid Earth}, 127\penalty0 (5):\penalty0 e2021JB023249, 2022.
\newblock \doi{https://doi.org/10.1029/2021JB023249}.

\bibitem[Zhu et~al.(2023)Zhu, Hou, Yang, Datta, Mousavi, Ellsworth, and Beroza]{zhu2023quakeflow}
Zhu, W., Hou, A.~B., Yang, R., Datta, A., Mousavi, S.~M., Ellsworth, W.~L., and Beroza, G.~C.
\newblock QuakeFlow: a scalable machine-learning-based earthquake monitoring workflow with cloud computing.
\newblock \emph{Geophysical Journal International}, 232\penalty0 (1):\penalty0 684--693, 2023.
\newblock \doi{https://doi.org/10.1093/gji/ggac355}.

\bibitem[Zhu et~al.(2025)Zhu, Wang, Rong, Yu, Zuzlewski, Tepp, Taira, Marty, Husker, and Allen]{zhu2025california}
Zhu, W., Wang, H., Rong, B., Yu, E., Zuzlewski, S., Tepp, G., Taira, T., Marty, J., Husker, A., and Allen, R.~M.
\newblock California Earthquake Dataset for Machine Learning and Cloud Computing, 2025.
\newblock \doi{https://doi.org/10.48550/arXiv.2502.11500}.

\end{thebibliography}

\includepdf[pages=-]{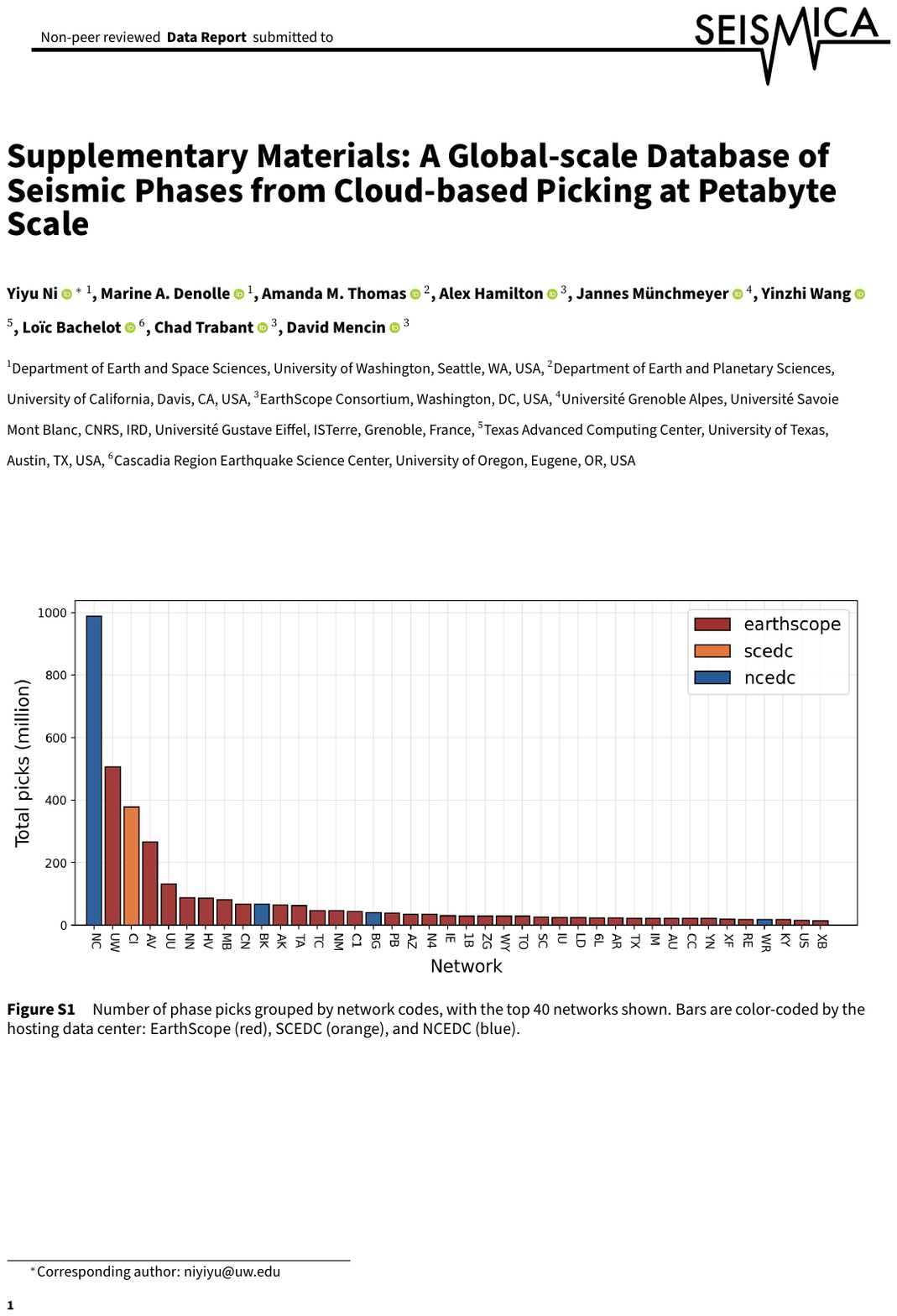}

\end{document}